\documentclass[a4,11pt]{article}

\usepackage{amsmath,amssymb,color,graphics,amscd,amsfonts,epsf,cite}
\allowdisplaybreaks[4]

\date{}
\begin{document}

\begin{flushright}

TAUP-2934/11
\end{flushright}

\vspace{0.1cm}

\begin{center}

{\LARGE
Generating new dualities through the  orbifold equivalence:

a demonstration in ABJM and four-dimensional quivers

  }
\end{center}
\vspace{0.1cm}
\vspace{0.1cm}
\begin{center}

         Masanori H{\sc anada}$^{a}$\footnote
         {
E-mail address : mhanada@u.washington.edu},
 Carlos H{\sc oyos}$^{a, b}$\footnote
         {
E-mail address : choyos@phys.washington.edu},
Andreas K{\sc arch}$^{a}$\footnote
         {
E-mail address : akarch@u.washington.edu}.

\vspace{0.3cm}

${}^a$ {\it Department of Physics, University of Washington,

 Seattle, WA 98195-1560, USA}\\

 ${}^{b}$
{\it Raymond and Beverly Sackler School of Physics and Astronomy

Tel-Aviv University, Ramat-Aviv 69978, Israel}\\

\end{center}

\vspace{1.5cm}

\begin{center}
  {\bf Abstract}
\end{center}

We show that the recently proposed large $N$ equivalence between ABJM theories with Chern-Simons terms of different rank and level, $U(N_1)_{k_1}\times U(N_1)_{-k_1}$ and  $U(N_2)_{k_2}\times U(N_2)_{-k_2}$, but the same value of $N^\prime =N_1 k_1=N_2 k_2$, can be explained using planar equivalence in the mirror duals. The combination of S-dualities and orbifold equivalence can be applied to other cases as well, with very appealing results. As an example we show that two different quiver theories with $k$ nodes can be easily shown to be Seiberg dual through the orbifold equivalence, but it requires order $k^2$ steps to give a proof when Seiberg duality is performed node by node.

\newpage

\section{Introduction and conclusions}

One of the most important insights about gauge theories gained over the last two decades is that gauge group and matter content aren't fundamental: different gauge theories, based on different groups, matter fields and couplings, can describe one and the same physical system. Large classes of such ``dualities" have been uncovered, many of them in supersymmetric gauge theories. Latter play an important role as they offer enough control over the dynamics of the theory to actually check a proposed duality. Examples of such dualities that we will consider include the exact S-duality \cite{Montonen:1977sn} of ${\cal N}=4$ supersymmetric gauge theories, mirror symmetry of three dimensional gauge theories \cite{Intriligator:1996ex} as well as Seiberg dualities \cite{Seiberg:1994pq} in three and four dimensions.

Presumably dualities are as omnipresent in non-supersymmetric gauge theories, but lacking the power of supersymmetry it is much harder to find and verify them. One notable exception are large $N$ orbifold equivalences \cite{Kachru:1998ys,Bershadsky:1998cb,Kovtun:2004bz,Lovelace:1982hz}, where a mother gauge theory is related to a daughter theory through a projection by a discrete subgroup of symmetries. Orbifold equivalence predicts that, in the limit of a large number of colors, the mother theory and the orbifolded daughter theory give identical physical observables (correlation functions, particle spectra), at least in a well defined subsector of the theories, neutral under the symmetry employed in the orbifold projection.

In the standard 't Hooft limit large $N$ equivalence can be proved diagrammatically to all orders in perturbation theory \cite{Bershadsky:1998cb} and has been shown to hold even non-perturbatively \cite{Kovtun:2004bz}, as long as certain symmetries intrinsic to the orbifolding procedure aren't dynamically broken. All these statements hold with or without supersymmetry.

Large $N$ equivalence between a supersymmetric mother and a daughter that may or may not be supersymmetric can be combined with well established dualities in supersymmetric gauge theories. This way novel dualities valid in the large $N$ limit can be deduced. This philosophy seems to have first been employed in \cite{Schmaltz:1998bg}, where a possible Seiberg dual pair of non-supersymmetric quiver gauge theories has been identified using large $N$ equivalence.\footnote{Other proposals for Seiberg dualities in non-supersymmetric theories can be found in \cite{seibergduals}.} While this duality is only derived at large $N$, it was observed in \cite{Schmaltz:1998bg} that at least the 't Hooft anomalies match between the two putative dual theories for all $N$, making this dual pair a candidate for an exact non-supersymmetric Seiberg dual. 

In this paper we will employ this basic strategy in two very interesting examples. The first concerns ${\cal N}=6$ Chern-Simons theory with gauge group $U(N)\times U(N)$ and levels $k$ and $-k$ introduced by Aharony, Bergman, Jafferis and Maldacena (ABJM) \cite{Aharony:2008ug} as the low energy effective theory of M2 branes on an orbifold.
It has recently been proposed in \cite{hhs} that for large $N$, finite $k$, a class of observables in the ABJM theory only depends on the product $Nk$.  That is, for a given $Nk$ they are completely independent of $k$.
This observation was based on the properties of its supergravity dual. From the field theory perspective, such independence of Chern-Simons theory on its level is rather unexpected. Here we will show that this statement can be understood as a consequence of mirror symmetry combined with a large $N$ orbifold equivalence. We can construct two ABJM theories with the same $Nk$ but different $k$ as the mirror dual of a mother and a daughter theory respectively. Interestingly, while the original mother and daughter pair are formulated in the standard planar 't Hooft large $N$ limit, the two equivalent ABJM theories are not: the 't Hooft limit would instruct us to work at fixed $N/k$, large $N$ (the type-IIA string theory limit of ABJM), whereas the orbifold equivalence holds at large $N$, fixed $k$ (the M-theory limit of ABJM).

Our second example involves a new Seiberg duality between two supersymmetric quiver theories. In contrast to the mirror symmetry case, Seiberg duality takes a theory in the planar limit into a dual theory that is also in the same limit. By orbifolding both the electric and magnetic theories of a known Seiberg dual pair, we obtain two daughters that are guaranteed to be equivalent at large $N$. The advantage of our supersymmetric example over the non-supersymmetric example of \cite{Schmaltz:1998bg} is that we can explicitly prove this duality by repeated node-wise application of a known Seiberg duality. In fact, we are able to show that the duality holds to {\it all orders} in $N$. Two lessons we learn from this are that a) the procedure of generating new duals from known duals using orbifold equivalences indeed works and that b) while the dualities are only derived in the large $N$ limit, the dual pairs obtained this way should be viewed as candidate duals that may be equivalent at all values of $N$. In particular, the fact that in our supersymmetric example large $N$ equivalence can be proved to hold to all orders in $N$ makes the observation of \cite{Schmaltz:1998bg} about the matching of anomalies to all orders in $N$ in the non-supersymmetric example even more intriguing. However, in order to prove that the necessary conditions stated in \cite{Kovtun:2004bz} hold in this case, a more detailed analysis is needed. For instance, it is known that in non-supersymmetric orbifolds of ${\cal N}=4$ super Yang-Mills, conformal invariance is broken at leading order in $N$ due to the running of double trace deformations \cite{tachyons}.

This paper is organized as follows. In the next section we review some basic properties of mirror symmetry and Seiberg dualities and in particular discuss their properties under taking large $N$ limits. In section 3 we present our mirror symmetry example, proving the large $N$ independence of level for ABJM Chern-Simons theory. In section 4 we discuss some simple examples of new Seiberg dualities that can be derived from old Seiberg dualities together with large $N$ equivalence and present our example of a supersymmetric quiver duality.
\section{Dualities and the large $N$ limit}
Dualities typically relate a strongly coupled theory to a weakly coupled theory. For example S-duality is the statement that an electric gauge theory with Yang-Mills coupling $g_{YM}$ has a dual description in terms of a magnetic gauge theory with coupling $1/g_{YM}$.
When discussing dualities in the large $N$ limit one has to therefore proceed with caution as typically the action of S-duality does not commute with the planar limit. In both the 't Hooft and the Veneziano limit one needs to take $N$ to infinity, $g_{YM}$ to zero, keeping the 't Hooft coupling $\lambda = g^2_{YM} N$ fixed (the difference between the two limits being whether the number of fundamental flavors, $N_f$, is kept finite or grows proportionally to $N$). Only this scaling allows one re-organize perturbation theory in terms of the genus of the Feynman diagrams. We'll therefore refer to either one of those two limits as a ``planar" limit. If the electric theory is taken to be in the planar limit, the S-dual magnetic theory clearly is not. Nevertheless, the magnetic theory may still simplify in the large $N$ limit (if it exists). In particular, orbifold equivalence plus S-duality predicts equivalences between two large $N$ theories that are taken in the $g^2_{YM} \rightarrow \infty$, $N \rightarrow \infty$ limit, which is definitely not a planar limit. One can, for example, start with ${\cal N}=4$ super-Yang-Mills with gauge group $U(N k)$ in the standard 't Hooft limit. Orbifolding by $\mathbb Z_k$ one can obtain a conformal daughter quiver theory \cite{Kachru:1998ys}, also in the `t Hooft limit (and ${\cal N}=2$, $1$ or $0$ supersymmetry, depending on how $\mathbb Z_k$ is embedded into the global $SU(4)_R$ symmetry of the ${\cal N}=4$ supersymmetric mother). Mother and daughter are equivalent at large $N$ in the sector neutral under $\mathbb Z_k$ symmetry. Among the properties the daughter inherits is conformality (at least at large $N$) and the exact S-duality\footnote{The action of S-duality in the daughter can be more complicated than a simple $SL(2,\mathbb Z)$ acting on the overall coupling. It also acts on the relative strength of the gauge couplings in the individual gauge groups as well as the individual $\theta$ angles, which are given by fluxes of NSNS and RR sector $B$ fields through the vanishing 2-cycles in the orbifold.} (as is clear from thinking about this in terms of D3 branes in IIB string theory located on a singular geometry). As a consequence, the magnetic duals of mother and daughter, if they exist, have to be equivalent. Many supersymmetric field theories can be embedded via brane constructions in type IIB string theory \cite{Hanany:1996ie}; the exact S-duality of the embedding type IIB string theory descends to an S-duality for the field theory on the branes and, as such, takes a theory in the planar limit to a theory that is not in the planar limit.

A different situation is encountered in Seiberg duality. Here, at least for the range of $N_f$ in the conformal window, both electric and magnetic theories can be taken in the planar limit. At low energies the field theory flows to a non-trivial fixed point with a fixed `t Hooft coupling $\lambda_*$ at which the beta function vanishes. Correspondingly, the way this duality arises from a string theory embedding \cite{Elitzur:1997fh} is very different in this case. Let us adapt the procedure of \cite{Elitzur:1997fh} to the case of Seiberg duality in 2+1 dimensions, so we can compare and contrast it with mirror symmetry. The brane setup defines a 3+1 dimensional gauge theory with 2+1 dimensional defects \cite{Karch:2000gx,DeWolfe:2001pq,Gaiotto:2008sa}, either defined on an interval or circle of length $L$. At energies below $1/L$ the 3+1 dimensional defect field theory flows to a 2+1 dimensional field theory which in turn flows to a strongly interacting 2+1 dimensional CFT. The matter content and gauge fields of this 2+1 dimensional field theory are determined by the matter living on the defects and the boundary conditions obeyed by the 3+1 dimensional fields on those defects; the only modes of the 3+1 dimensional fields that survive are the zero modes, that is modes independent of the coordinate along the compact dimension. So one would expect that rearranging the order and relative positions of the defects along the internal direction is an irrelevant deformation and all such theories will flow to the same IR CFT. However, under certain such brane moves the naive 2+1 dimensional field theory one associates to these branes changes from the original electric theory to its ``magnetic" Seiberg dual. While the 3d gauge couplings associated with the individual 2+1 dimensional gauge group factors are set by the sizes of the intervals,  none of these operations acts on the 3+1 dimensional gauge coupling, which is set by the string coupling $g_s$. So this whole discussion can be entirely kept within the planar limit of the 3+1 dimensional defect CFT.

Somewhat in contrast, mirror symmetry is inherited from the S-duality of the embedding IIB field theory. As such one would expect it to relate a theory in the planar limit to a non-planar theory. Strictly speaking, this S-duality is a property of the 3+1 dimensional defect CFT, a point most clearly made in \cite{Gaiotto:2008ak}. A 3+1 dimensional defect theory with coupling $g^2_{YM}$ on a circle of radius $L$ is dual to a different 3+1 dimensional defect theory with coupling $1/g^2_{YM}$ on a circle of the same radius $L$. To obtain from this a statement about 2+1 dimensional theories, we want to study the theory at scales below $1/L$. If the original 3+1 dimensional theory is in the planar limit, so will be the dimensionally reduced 2+1 theory with a 't Hooft coupling $\lambda_{3d} = g^2_{YM} N/L$. The S-dual however will not be, and in this case it is not even clear how to read off a 2+1 gauge theory. However, appealing once more to the irrelevance of the positions of the defects on the compact direction one can deform both brane setups into a regime where one can associate a 2+1 dimensional weakly coupled gauge theory with them at scale $1/L$ and then appeal to universality to argue that they will flow to the same IR CFT as the original S-dual pair.
Although it is difficult to justify this procedure in mathematically robust manner, in the example in \S\ref{sec: ABJM}, it reproduces a nontrivial equivalence
which is expected from the $AdS_4/CFT_3$ correspondence.  We regard it as a justification of the procedure.
\section{ABJM orbifold equivalence}\label{sec: ABJM}
\subsection{ABJM and orbifold equivalence in the gravity dual}

The field content of ABJM consists of two ${\cal N}=2$ $U(N)$ vector multiplets, an adjoint chiral multiplet for each gauge group and four chiral multiplets in the bifundamental representation. The action for the gauge fields is a supersymmetric version of the Chern-Simons action, with opposite levels for each gauge group. In components,
\begin{equation}
S_{CS}=\frac{k}{4\pi}\int d^3 x\,{\rm Tr}\,\left\{
\epsilon^{\mu\lambda\nu}\left(
A_\mu\partial_\lambda A_\nu+\frac{2}{3}A_\mu A_\lambda A_\nu
- \hat A_\mu\partial_\lambda \hat A_\nu-\frac{2}{3}\hat A_\mu \hat A_\lambda \hat A_\nu
\right)
-\overline{\chi}\chi+2D\sigma+\overline{\hat \chi}\hat \chi-2\hat D\hat \sigma
\right\}.
\end{equation}
The ${\cal N}=2$ vector multiplet can be obtained from the dimensional reduction of an ${\cal N}=1$ vector multiplet in four dimensions. Then, the scalar fields $\sigma$, $\hat \sigma$ can be seen as the 4th components of the reduced gauge fields. In addition, each multiplet includes a Dirac spinor $\chi$, $\hat \chi$ and an auxiliary scalar $D$, $\hat D$. Under large gauge transformations the Chern-Simons action is not invariant, but it would introduce a phase in the path integral. Fixing the level $k$ to be an integer makes this phase trivial. The adjoint chiral multiplets have a quadratic superpotential proportional to the Chern-Simons level, and couple to the bifundamental fields through a cubic superpotential. After integrating them out, they produce a marginal quartic superpotential. Then, chiral multiplets have a sextic potential, with additional contributions that appear after integrating out the fields $\sigma$ and $\hat \sigma$. It is possible to show that the theory has an enhanced $SU(4)_R\simeq SO(6)_R$ symmetry, and an explicit ${\cal N}=6$ supersymmetry. For $k=1$ and $k=2$ the symmetry is further enhanced to ${\cal N}=8$, with the additional $SO(8)_R$ currents appearing thanks to monopole operators.

The $k=1$ ABJM theory is the only ${\cal N}=8$ superconformal theory in three dimensions that is a candidate to describe low-energy physics of a stack of $N$ M2 branes in flat spacetime\footnote{The original proposal \cite{Bagger:2007jr} for $N=2$ was shown to be equivalent to ABJM with $SU(2)\times SU(2)$ gauge group.}. Given the amount of supersymmetry, it may be unique. For $k> 1$, it is believed to describe M2 branes in an orbifold  $\mathbb{R}^8/\mathbb{Z}_k$, with ${\cal N}=6$ supersymmetry when $k>2$. In the large-$N$ limit with $N\gg k$ (actually $N\gg k^5$), the ABJM theory should be dual to the near-horizon geometry of the M2 branes, M-theory on $AdS_4\times S^7/\mathbb{Z}_k$. This has been confirmed by many calculations comparing both theories. For large values of $k$ ($k^5\gg N \gg k$), the M-theoretic description is not a good approximation but a weakly coupled description in terms of type IIA supergravity in $AdS_4\times \mathbb{CP}^3$ is possible.

The metric and background fields of the gravity dual to ABJM are
\begin{eqnarray}
\notag &ds^2=\frac{R^2}{4} ds^2_{AdS^4}+R^2 ds^2_{S^7/\mathbb{Z}_k}, \\
&F_4 \sim N' \epsilon_4, \\
\notag &R/\ell_P =(2^5 \pi^2 N')^{1/6}.\label{11d_curvature}
\end{eqnarray}
The metric $ds^2_{AdS_4}$ is the usual anti-de Sitter metric with unit radius. The $S^7$ geometry can be seen as a fibration of a circle on $\mathbb{CP}^3$, with the orbifold acting on the circle. This reduces the volume of the $S^7$ by a factor of $k$, so the number of flux quanta is also reduced by a factor of $k$ in the quotient space and one needs $N'=Nk$ units of seven form-flux (the magnetic dual to the four-form flux) to match with the flux of $N$ M2 branes. Notice that the radius of the M2 geometry depends also on $N'$, and not on $N$.

Consider now two different ABJM theories, with ranks $N_1$ and $N_2$ and levels $k_1$ and $k_2$. As was pointed out in \cite{hhs}, if $N_1 k_1=N_2 k_2=N'$, the gravity duals to both theories would have the same flux and radius of curvature, although the orbifold would be different. From the point of view of supergravity, this is completely analogous to the orbifold projections that have been considered in the context of $AdS_5/CFT_4$ duality~\cite{Kachru:1998ys}. On the gravity side, by starting with type IIB superstring on $AdS_5\times S^5$, one can perform the orbifold projection on $S^5$ so that type IIB superstring on $AdS_5\times S^5/{\mathbb Z}_k$ is obtained. The dual conformal field theory, which was originally  ${\cal N}=4$ super Yang-Mills in four dimensions, is projected to less symmetric theories.
(This procedure on the gauge theory side is also called the `orbifold projection'.) It was found using the gravity dual that one can predict an equivalence between the corresponding field theories, for observables
which are invariant under ${\mathbb Z}_k$ projection.

The argument so far is completely parallel to the one used in $AdS_5/CFT_4$ duals.\footnote{There is a technical but important difference between the ${\mathbb Z}_k$ orbifolds in $AdS_5/CFT_4$ duals and the ones considered here. In the $AdS_5/CFT_4$ setup the orbifold acts non-freely, i.e. there is a subspace of fixed points where light states are localized. From the field theory perspective large $N$ orbifold equivalence does not apply to observables in this sector and can be spoiled if they acquire an expectation value. The difference is that in the M-theory geometry the ${\mathbb Z}_k$ orbifold acts freely on the $S^7$, so there are no fixed points and therefore no such localized light states.} However, from the gauge theory point of view there is a big difference. In $AdS_5/CFT_4$ duals, it is possible to prove the equivalence directly in the gauge theory~\cite{Bershadsky:1998cb,Kovtun:2004bz,Lovelace:1982hz}; the essential point in the proof is that loop effects in the gravity side correspond to $1/N$ corrections in the gauge theory side. Hence one should focus in the planar limit, and the equivalence is actually proven there, using purely field theory arguments.
However, in our case the same kind of arguments are not applicable. The reason is as follow: in ABJM theories, the 't Hooft expansion is the double expansion in terms of $1/N$ and the 't Hooft coupling $\lambda=N/k$. Usual 't Hooft counting holds in the large-$N$ limit with fixed $\lambda$, which is dual to type IIA superstring on $AdS_4\times {\mathbb CP }^3$~\cite{Aharony:2008ug}. In the parameter region of interest, however, $k$ is typically of order unity and  $N$ is taken to be very large, so that
the dominance of the planar diagrams cannot be used.\footnote{
Of course, one can consider the more familiar type of 
orbifold equivalences associated with planar diagrams in the 
$AdS_4/CFT_3$ context as well, but in the limit where $N/k$ is fixed,
not in the M-theoretic limit we are interested in. 
See comments in \cite{hhs} and refs. \cite{Benna:2008zy}. }
Thus the $AdS_4/CFT_3$ duality, combined with the orbifold equivalence,
predicts non-trivial relations for very strongly coupled
(in the sense that the 't Hooft expansion cannot be used) ABJM theories.
In the following we will give a proof of the equivalence using mirror duality.

\subsection{Mirror dualities  plus orbifold equivalence}

We can explain the large-$N$ equivalence among ABJM theories with the same $N'=Nk$ using a combination of mirror symmetry \cite{Intriligator:1996ex} and orbifold equivalences \cite{Kachru:1998ys,Bershadsky:1998cb,Kovtun:2004bz,Lovelace:1982hz}. The particular examples of mirror symmetries in ABJM-like theories we will need here were studied in ref. \cite{Jensen:2009xh}. Let us first recall how to obtain the ABJM theory from string theory. One starts with two parallel NS5 branes extended along the 012345 directions and separated along the compact 6 direction, and $N$ D3 branes along the 0126 directions. In addition, there are $k$ D5 branes along the 012349 directions. This leads to a low energy effective theory on the worldvolume of D3 branes that is an ${\cal N}=2$ supersymmetric theory in three dimensions with $U(N)\times U(N)$ gauge group, a chiral multiplet in the adjoint representation of each gauge group factor ($X_i$, $i=1,2$), four bifundamental chiral multiplets ($A$, $\widetilde A$, $B$, $\widetilde B$) and $k$ massless chiral multiplets in the fundamental representation ($Q_{ai}$) and $k$ in the anti-fundamental ($\widetilde Q_{ai}$) of each gauge group factor $U(N)$. If the $k$ D5 branes sit on top of one of the NS5 branes, one can deform the brane configuration to a $(1,k)$5-brane,\footnote{The $(1,k)$5 brane is at an angle on the 59 plane.} that in the effective field theory corresponds to giving masses of opposite sign to chiral multiplets charged under each group\cite{Bergman:1999na}. After integrating out those, one is left with Chern-Simons terms for the gauge fields with level $k$ and level $-k$ for each gauge group factor. 
This is a marginal term, while Maxwell's action for the gauge fields is irrelevant in 2+1 dimensions, in such a way that at low energies the theory flows to an ${\cal N}=6$ superconformal theory, which is enhanced to ${\cal N}=8$ when $k=1$ or $k=2$.

\begin{figure}[htbp]
\begin{center}
\scalebox{0.33}{
\rotatebox{-90}{
\includegraphics{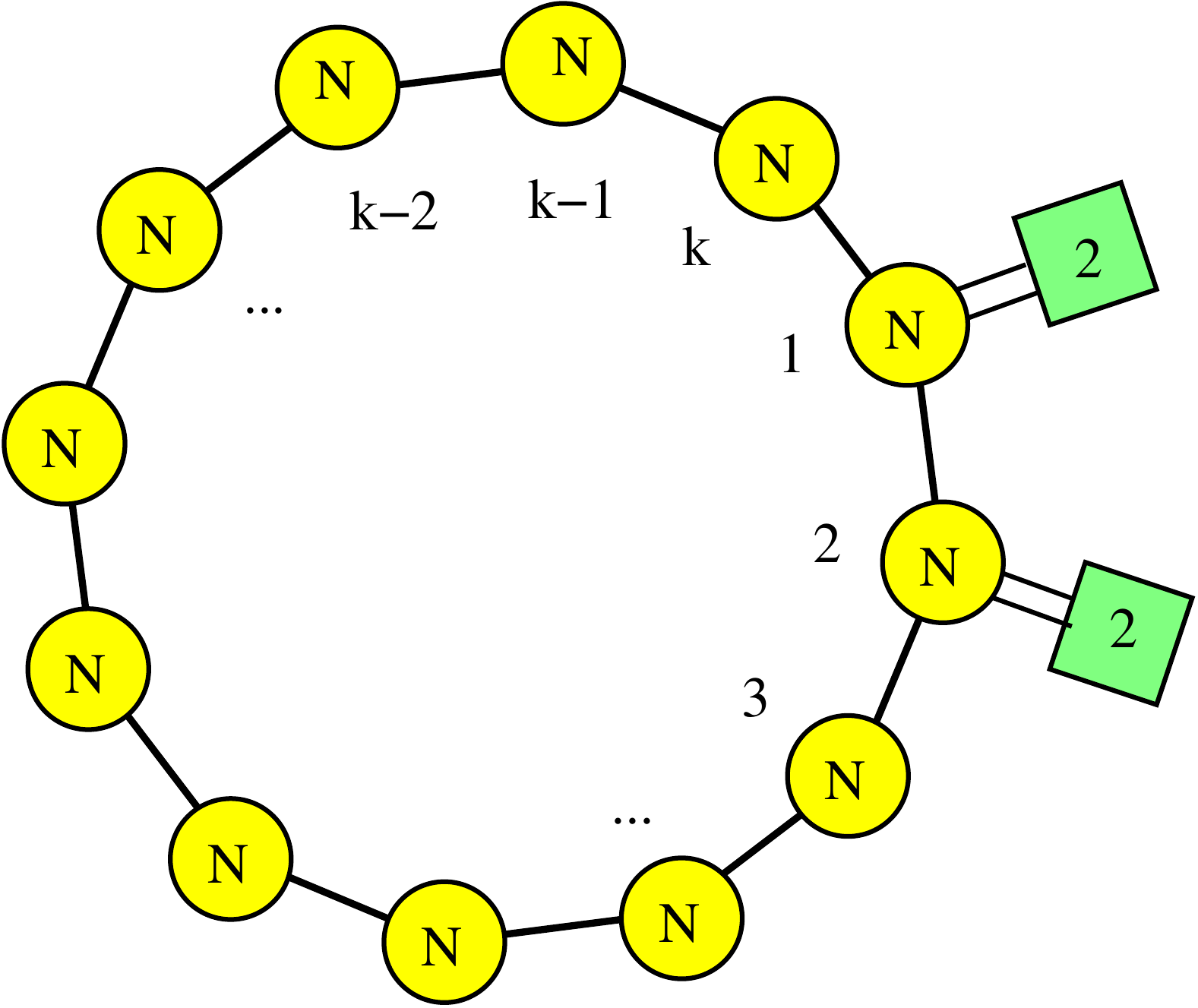}}}
\caption{
Quiver of the S-dual theory to the ABJM construction. Each node represents a $U(N)$ vector multiplet plus an adjoint chiral multiplet, each link a bifundamental hypermultiplet and each box a global flavor group.
}
\label{fig:quiver}
\end{center}
\end{figure}

Mirror duality \cite{Intriligator:1996ex} can be studied using the exact S-duality transformations of the string theory construction. Under S-duality, the NS5 branes and the D5 branes are interchanged, while the D3 branes are self-dual. Before adding a mass to the fundamental fields, the S-dual of the ABJM construction consists of $k$ NS5 branes along the 012578 directions intersecting a D5 brane along the 012345 directions, with an additional D5 separated along the 6 direction, and the same $N$ D3 branes along 0126 directions \cite{Jensen:2009xh}. The effective theory on the D3 branes is an ${\cal N}=2$ $U(N)^k$ gauge theory. If we label each gauge group by a number $i=1,2,\dots,k$, the matter content consists of $k$ chiral multiplets in the adjoint ($Y_i$), $2k$ chiral multiplets the bifundamental representation of $(i,i+1)$ groups \footnote{Identifying the $k+1$th group with the $1$st.} ($C_{i\,i+1}$, $\widetilde C_{i+1\,i}$), one chiral multiplet in the fundamental representation and one in the anti-fundamental for each the $1$st and the $2$nd gauge group factors ($R_{i}$, $\widetilde R_{i}$, $i=1,2$). We can summarize the field content as a quiver, like the one in figure \ref{fig:quiver}. The superpotential is \cite{Brunner:1998jr}
\begin{equation}\label{Wquiver}
W\propto \widetilde R_1 C_{12} R_2-\widetilde R_2 \widetilde C_{21} R_1,
\end{equation}
so the flavor symmetry is reduced from $U(1)^2\times U(1)^2$ to $U(1)\times U(1)$. The deformation to the $(1,k)$5-brane that was a mass for the fundamental multiplets in the original theory becomes a complicated non-local deformation in the mirror. However, due to the duality, both should have the same physical effect and the two theories should flow to the same IR fixed point.

A quiver theory of the form of the mirror dual is very similar to the theories studied in the usual orbifold projections \cite{Kachru:1998ys,Bershadsky:1998cb}. Indeed, we can obtain the same field theory by doing a $\mathbb{Z}_k$ orbifold projection of an ${\cal N}=2$ $U(kN)$ gauge theory with three chiral multiplets in the adjoint representation ($Y$, $C$, $\widetilde C$), two chiral multiplets in the fundamental representation ($R_a$, $a=1,2$) and two in the anti-fundamental ($\widetilde R_a$).The superpotential takes the form
\begin{equation}
W \propto \widetilde R_1 C R_2-\widetilde R_2 \widetilde C R_1,
\end{equation}
so there is only a $U(1)\times U(1)$ global flavor symmetry also for this theory. We will now show explicitly the orbifold projection. We start defining the phase $\omega=e^{2\pi i/k}$. The projection is made using an element of the gauge group $\gamma\in U(kN)$ that spans a $\mathbb{Z}_k$ subgroup:
\begin{equation}
\gamma={\rm diag}\, \left(\;\mathbf{1}_N,\; \omega\mathbf{1}_N,\; \omega^2\mathbf{1}_N,\; \dots\; ,\; \omega^{k-1}\mathbf{1}_N\; \right)
\end{equation}
where $\mathbf{1}_N$ is the $N\times N$ identity matrix. The quiver theory is obtained from the $U(kN)$ theory by keeping the components of the fields that are invariant under the following transformations:
\begin{itemize}
\item Fields in the vector multiplet: $V\to \gamma V \gamma^{-1}$.
\item Adjoint chiral multiplets: $Y\to \gamma Y \gamma^{-1}$, $C\to \omega \gamma C \gamma^{-1}$, $\widetilde C\to \omega^{-1}\gamma \widetilde C \gamma^{-1}$ .
\item Fundamental chiral multiplets: $R_1\to \gamma R_1$, $R_2\to \omega^{-1}\gamma R_2$.
\item Anti-fundamental chiral multiplets: $\widetilde R_1\to  \widetilde R_1 \gamma^{-1}$  $\widetilde R_2\to  \omega\widetilde R_2 \gamma^{-1}$.
\end{itemize}
The transformations we have used are a combination of gauge, flavor and R-symmetry transformations that are compatible with the form of the superpotential. After the orbifold projection, the $U(N)^k$ theory inherits the superpotential \eqref{Wquiver}.

\begin{figure}[htbp]
\begin{center}
\scalebox{0.45}{
\rotatebox{-90}{
\includegraphics{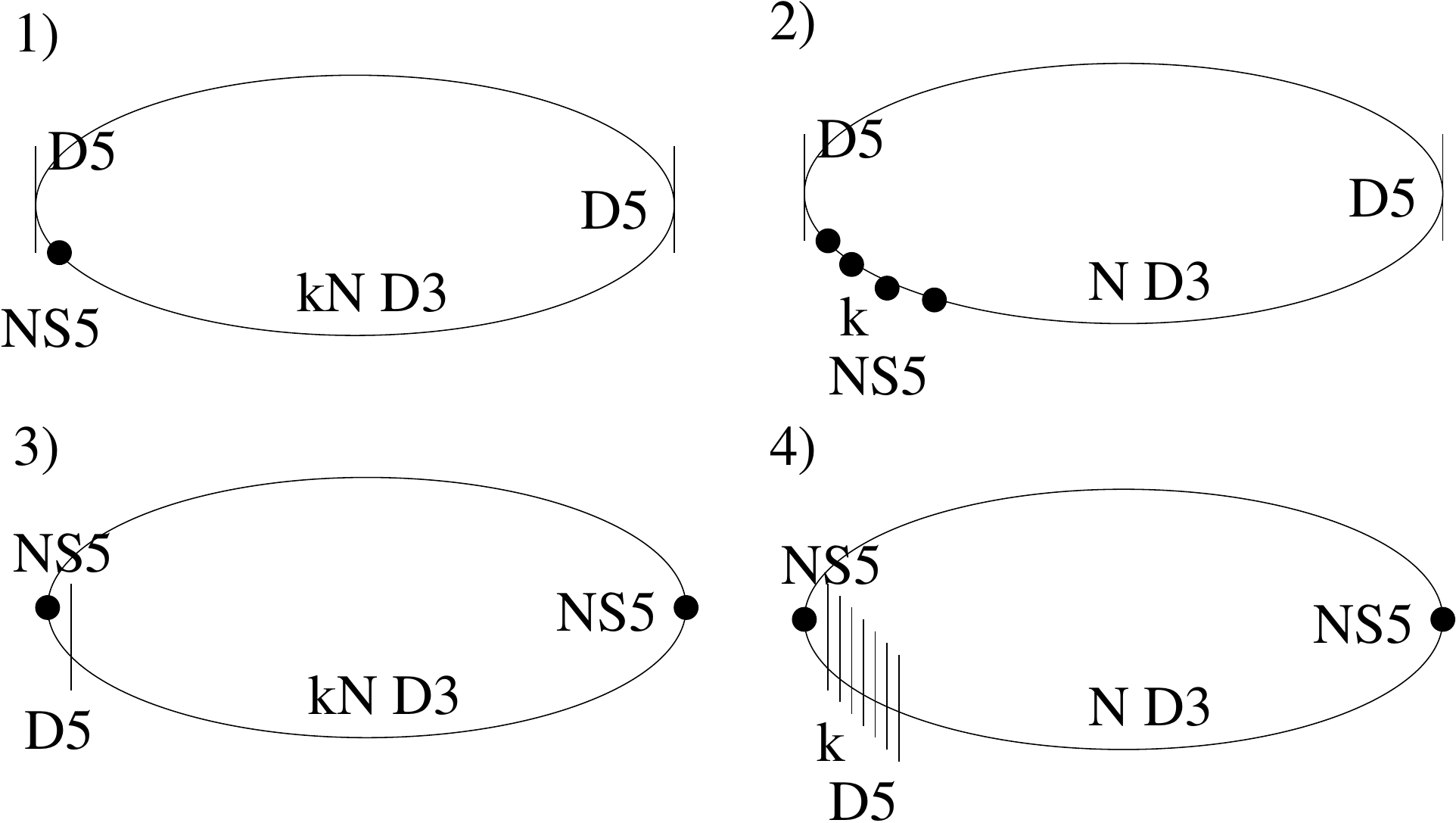}}}
\caption{
D-brane constructions for the different theories discussed in the text. The circle represents D3 branes wrapping a compact direction, while thick dots represent NS5 branes and vertical lines represent D5 branes. Although NS5 branes and D5 branes that are on the same side of the circle should be on top of each other, we have separated them in the drawing for visual convenience. At low energies, the effective field theory on the D3 branes is 1) $U(kN)$ with 4 flavors,  2) $U(N)^k$ with 4 flavors, 3) $U(kN)\times U(kN)$ with 2 flavors, 4) $U(N)\times U(N)$ with $2k$ flavors. The pairs 1), 3) and 2), 4) are related by S-duality, while the pair 1), 2) is related by the orbifold equivalence.
}\label{fig:dbranes}
\end{center}
\end{figure}

The $U(kN)$ theory can also be obtained from a brane configuration, it consists of $kN$ D3 branes along the 0126 directions, one NS5 brane along the 012578 directions and two D5 branes along the 012345 directions. An S-duality gives a D5 brane along the 012349 directions and two NS5 branes along the 012345 directions, so at low energies one obtains a $U(kN)\times U(kN)$ theory of the kind discussed at the beginning of this section, with one chiral multiplet in the fundamental representation and one in the anti-fundamental for each gauge group. Giving opposite mass to the chiral multiplets charged under each gauge group, the theory becomes the ABJM theory $U(kN)_1\times U(kN)_{-1}$ in the infrared. We have then the following chain of relations
\begin{equation}
\begin{array}{ccc}
U(kN)\,+\,4\; {\rm flavors} & \overset{\rm orbifold}{\longleftrightarrow} &  U(N)^k\,+\,4\; {\rm flavors}\\
{\rm mirror}\,{\Big\updownarrow} & & {\rm mirror}\,{\Big\updownarrow} \\
 U(kN)\times U(kN)+\,2\; {\rm flavors} &  & U(N)\times U(N)\,+\,2k\;  {\rm flavors}
\end{array}
\end{equation}
We have represented the corresponding D-brane constructions in figure \ref{fig:dbranes}.

Let us discuss the validity of our derivation: we start with a $U(N)\times U(N)$ theory with $2k$ massless flavors, this theory flows in the IR to a fixed point whose mirror is given by a $U(N)^k$ theory with four flavors. Mirror duality in the IR is assumed to be exact by virtue of S-duality in string theory. The mirror theory is orbifold equivalent to a $U(kN)$ theory with four flavors. In order for the orbifold equivalence to be valid, both the $U(N)^k$ and the $U(kN)$ theory need to be in the 't Hooft limit, this imposes some restrictions. As mentioned above, these theories can be seen as the dimensional reduction of four-dimensional theories, with dimensionless coupling constant $g_{YM}^2$. The 't Hooft coupling of the three dimensional theory is determined by the size of the compact direction $L$
\begin{equation}
\lambda_{3d}=\frac{g_{YM}^2 N}{L}.
\end{equation}
Assuming the four-dimensional theory is in the 't Hooft limit, $N\to \infty$ with $g_{YM}^2N$ fixed, but possibly large, the three-dimensional theory would also be in the 't Hooft limit unless non-trivial dynamics happen at very low energy scales $E\sim g_{YM}^2/L\sim 1/(NL)$, where a separate expansion in $N$ and the effective coupling $\lambda_{3d}/E$ would break down. We do not expect this to be the case, rather the two theories will flow to a fixed point at scales below $\lambda_{3d}$, and they will be related by the orbifold equivalence, that has been shown to hold at the non-perturbative level \cite{Kovtun:2004bz,Unsal:2010qh}. However, in general this is a subtle question and non-trivial dynamics can be relevant at very low energies if for instance chiral symmetry is broken \cite{Unsal:2010qh}. Once the equivalence between the $U(N)^k$ theory and the $U(kN)$ theory has been established, one can apply mirror duality to the latter and end with the $U(kN)\times U(kN)$ theory with 2 flavors.
Finally we have to introduce a mass deformation for the fundamental fields in the $U(N)\times U(N)$ theory. The theory flows to a new fixed point described by ABJM. The mass deformation maps to a relevant, possibly non-local, deformation in the mirror dual $U(N)^k$. Giving the same mass to all the flavors, makes the deformation invariant under $\mathbb{Z}_k$ symmetry, so the deformation is mapped by the orbifold equivalence to a deformation in the $U(kN)$ theory, and through mirror duality to a mass for flavors in the $U(kN)\times U(kN)$ theory. Therefore, there is an equivalence between the $U(kN)_1\times U(kN)_{-1}$ and the $U(N)_k\times U(N)_{-k}$ ABJM theories. Since we have done this for arbitrary $k$ and $N$, this means that there is an equivalence between any two ABJM theories with ranks $N_1$, $N_2$ and levels $k_1$, $k_2$, such that
\begin{equation}
N'=N_1k_1=N_2 k_2,
\end{equation}
are the same.

\section{Orbifolds of Seiberg duals and Seiberg duals of orbifolds}
\subsection{Simple examples}
The idea of combining orbifold equivalences with dualities can be applied to other cases as well. One of the best studied type of duality is Seiberg duality between two four-dimensional ${\cal N}=1$ supersymmetric gauge theories \cite{Seiberg:1994pq}. Seiberg duality is a kind of electromagnetic duality where two different theories, the ``electric'' and the ``magnetic'' theory, flow in the infrared to the same fixed point, which is usually strongly coupled except for some values of the parameters. The canonical example\footnote{In 3+1 dimensions one typically only considers $SU(N_c)$ gauge groups as the $U(1)$ factor in $U(N_c)$ becomes free in the infrared and so trivially agrees on both sides. Seiberg duality can however also be studied in 2+1 dimensions. It has been shown in  \cite{Karch:1997ux,Aharony:1997gp} that the derivation of Seiberg duality in terms of irrelevant positions of defects equally applies in 2+1 and 3+1 dimensions. The duality for $U(N_c)$ gauge group with $N_f$ fundamentals has been proposed in \cite{Aharony:1997gp} in close analogy with the 3+1 dimensional case. In 2+1 dimensions it is crucial to keep the $U(1)$ factor; the $U(1)$ here is strongly interacting in the infrared and so the dynamics of $U(N_c)$ truly differs from that of $SU(N_c)$. While our discussion is focused on the 3+1 dimensional case, we keep considering $U(N_c)$ gauge groups as generalization to 2+1 dimensions in this case proceeds without any major changes.} is $U(N_c)$ SQCD with $N_f$ flavors in the conformal window $\frac{3}{2} N_c < N_f < 3 N_c$. The dual theory is $U(N_f-N_c)$ SQCD $N_f$ flavors and additional ``meson'' multiplets.

Taking $N_c=k n_c$, we can perform a $\mathbb Z_k$ orbifold to obtain a $U(n_c)^k$ quiver gauge theory. In order for the $k$ nodes in this quiver to not completely decouple from each other, we need bi-fundamental matter. In orbifolding ${\cal N}=4$ this bi-fundamental matter is the result of adjoint representation matter that is explicitly charged under the $\mathbb Z_k$. The embedding of $\mathbb Z_k$ into the gauge group keeps only the $U(n_c)$ blocks on the diagonal of the gauge field invariant. If the adjoint matter only transforms under $\mathbb Z_k$ due to its gauge charges, the same components as for the gauge field are invariant and we will project down to $k$ decoupled $U(n_c)$ factors with adjoint matter. If the adjoint has some additional explicit $\mathbb Z_k$ charge, it is the off-diagonal blocks that survive the projection and we are left with bi-fundamental matter.

In the basic Seiberg duality described above the only adjoint matter are the gauginos. If we want to preserve ${\cal N}=1$ supersymmetry the gauginos have to transform the same way under $\mathbb Z_k$ as the gauge fields and the only option we have is to project down to $k$ decoupled $U(n_c)$ gauge groups with $n_f$ flavors each (here $N_f=k n_f$), and similarly to $k$ decoupled $U(n_f - n_c)$ factors in the magnetic dual. The orbifold daughters are trivially Seiberg dual to each other -- they are just $k$ copies of a standard Seiberg dual pair. In \cite{Schmaltz:1998bg} $\mathbb Z_k$ was embedded non-trivially into the global $U(1)_R$ symmetry. This way the gaugino gave rise to chiral bi-fundamental matter after projection at the cost of losing supersymmetry. In the next section we will instead introduce an extra adjoint chiral multiplet into the theory. Seiberg duality in this case is still well understood, and so we can produce interesting supersymmetric Seiberg dual quivers.

There is one non-trivial Seiberg dual we can derive from the simple example of $U(N_c)$ with $N_f$ flavors by performing a $\mathbb Z_2$ projection on the gauge group that takes $SO(2N_c)$ with $N_f$ flavors to $U(N_c)$ with $N_f$ flavors \cite{Cherman:2010jj,Hanada:2011ju}. The dual gets projected to $SO(N_f-N_c)$. $SO(N_c)$ gauge groups are known to have a $SO(N_f-N_c+4)$ magnetic dual \cite{Intriligator:1995id}. At large $N_c$ the 4 is negligible compared to $N_f$ and $N_c$, so this large $N$ orientifold equivalence indeed produced a valid dual pair, as long as we stay away from the edges of the conformal window where the symmetry breaking patterns in mother and daughter do not necessarily agree. Note that in this case the duality between the daughters does {\it not} extend to finite $N$, in contrast to the case we are about to turn to.
\subsection{A quiver Seiberg duality}
Seiberg duality can be extended to more complicated theories. In \cite{Kutasov:1995ve,Kutasov:1995np} an extension of Seiberg duality to SQCD theories with an additional adjoint chiral multiplet $X$ and a superpotential $W(X)=  {\rm Tr}X^k$  ($k\geq 3$) was studied. The dual is $U((k-1)N_f-N_c)$ SQCD with an adjoint multiplet $Y$ and a superpotential of the form
\begin{equation}
W(Y)=  {\rm Tr}Y^k+\sum_{j=1}^{k-1} M_j \tilde{q} Y^{k-j-1} q,
\end{equation}
where $q$, $\tilde{q}$ are the dual quark multiplets and $M_j$ the meson multiplets dual to composite operators $\tilde{Q} X^{j-1} Q$ in the electric theory. Note that the F-term associated with the $X^k$ superpotential in the electric theory set to zero $X^{k-1}$ and so the chiral ring truncates and there are no mesons of this form with $j>k-1$ in the spectrum. We will consider a particular family of examples of this type, with gauge group $U(k(k-1)n_c)$, and $k n_f$ flavors, whose Seiberg dual has a gauge group $U(k(k-1)(n_f-n_c))$.
Note that we chose the same integer $k$ to determine the gauge group as well as the superpotential.
The reason behind this choice is that we can do the same $\mathbb{Z}_k$ orbifold projection in both the electric and the magnetic theory. The $\mathbb Z_k$ is embedded in the gauge group in the usual fashion, projecting the electric theory down to $U( (k-1) n_c)^k$ and similarly the magnetic gauge theory down to $U( (k-1) (n_f - n_c))$. In order to obtain a non-trivial quiver with bi-fundamental matter, we also need to act on the adjoint chiral fields $X$ and $Y$ explicitly. In the absence of any superpotentials, we could use the $U(1)$ global symmetry that rotates the chiral superfield $X$ while leaving all other fields invariant and the analogous symmetry acting on $Y$ in the magnetic theory. Note that our particular choice of superpotential leaves a $\mathbb{Z}_k$ subgroup of this $U(1)$ global symmetry invariant and so can indeed be used for a projection.

\begin{figure}[!htbp]
\begin{center}
\scalebox{0.35}{
\includegraphics{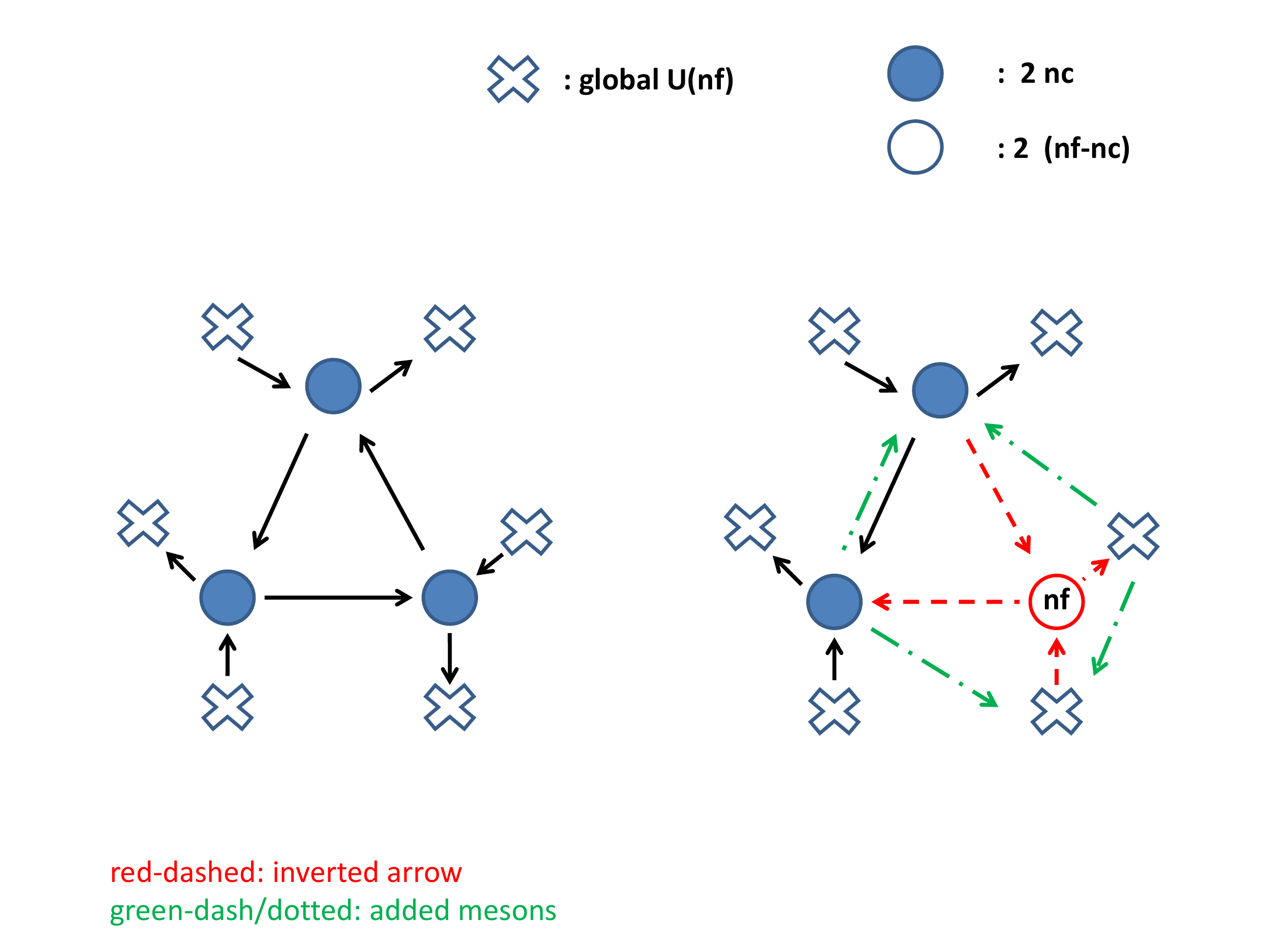}}
\caption{
First step in the dualization for $k=3$. The algorithm is applied to the first node. Reverted arrows and added mesons are indicated as well as the dualized gauge group. The new quiver can be simplified by integrating out massive mesons, as indicated in the starting point for the next step in the following figure.
\label{k3one}
}
\end{center}
\end{figure}

The field content of the resulting theories can be summarized in a circular quiver with two flavor groups attached to each node, as the left diagram in figure \ref{k3one}. In this kind of quiver links $F_i$ are oriented, since we are dealing with ${\cal N}=1$ chiral multiplets. Each quiver has $k$ nodes corresponding to an $U((k-1)n_c)^k$ group in the orbifold of the electric theory and a $U((k-1)(n_f-n_c))^k$ group in the orbifold of the magnetic theory. The links joining the nodes correspond to the bifundamental fields obtained after projecting the adjoint multiplets $X$ and $Y$. In addition both theories have a superpotential of the form $\prod_{i=1}^k F_i$ that descends from the $X^k$ and $Y^k$ superpotentials in the parents. This is consistent with the expectation that in quiver theories one wants to include a superpotential term associated with every closed loop. The electric and magnetic quiver look identical as far as the nodes and the charged matter is concerned except for the change of electric to magnetic gauge group on every node. In addition the magnetic theory has a large set of mesons. For every global flavor group associated with a quark we have a meson arrow going from that flavor group to every single flavor group associated to an anti-quark. In the electric theory these mesons are operators of the form $Q \left ( \prod_{i=1}^{j+1} F_i \right ) \tilde{Q}$, that is we take $j+1$ link fields to go partway around the quiver and then make a gauge invariant by attaching a flavor field at the start and end node.

In contrast to the mirror duals, for both the electric and the magnetic theory the planar approximation is valid in the large-$N_c$ limit, as long as they are not too close to the border of the conformal window, where the coupling of one of the two becomes very weak and the other very strong. We can then apply the usual argument of planar equivalence in both cases, and since the Seiberg duals flow to the same infrared fixed point, this means that the orbifolded theories should also flow to the same fixed point, which differs from the original one by $1/N_c$ corrections\footnote{Strictly
speaking large-$N_c$ equivalence in the Veneziano limit $N_f/N_c\sim 1$ has not been proven at nonperturbative level,
although a perturbative proof \cite{Bershadsky:1998cb} can easily be applied to this case \cite{Hanada:2011ju}.
However, it should be possible to apply the arguments for adjoints and quiver theories as in \cite{Kovtun:2004bz}
just by replacing one of the gauge groups with the flavor group. In this case the equivalence is expected to hold only for flavor singlet observables.
}.
In principle one should be able to implement the orbifold projection directly at the fixed point. In this case, we are doing the same orbifold projection to one and the same theory and so the orbifolded theories would have to be Seiberg dual to each other including $1/N_c$ corrections eve in the non-supersymmetric case studied in \cite{Schmaltz:1998bg}. However in practice we implement the orbifold projection in the UV, and in this case $1/N_c$ corrections could drive the ``electric'' and/or the ``magnetic'' theories away from the fixed point. In this case they would not flow to the same theory.

Can we test our claim? The procedure to find the Seiberg dual of a quiver theory has been explained in \cite{Franco:2005rj}. Seiberg duality can be applied to each gauge group factor independently, treating the remaining gauge groups as global flavor groups. Clearly, for the quivers we consider, it is not enough to apply Seiberg duality to a single gauge group. We have found a simple algorithm to derive the Seiberg dual we have obtained through the orbifold projection using this procedure.

\begin{figure}[!htbp]
\begin{center}
\scalebox{0.35}{
\includegraphics{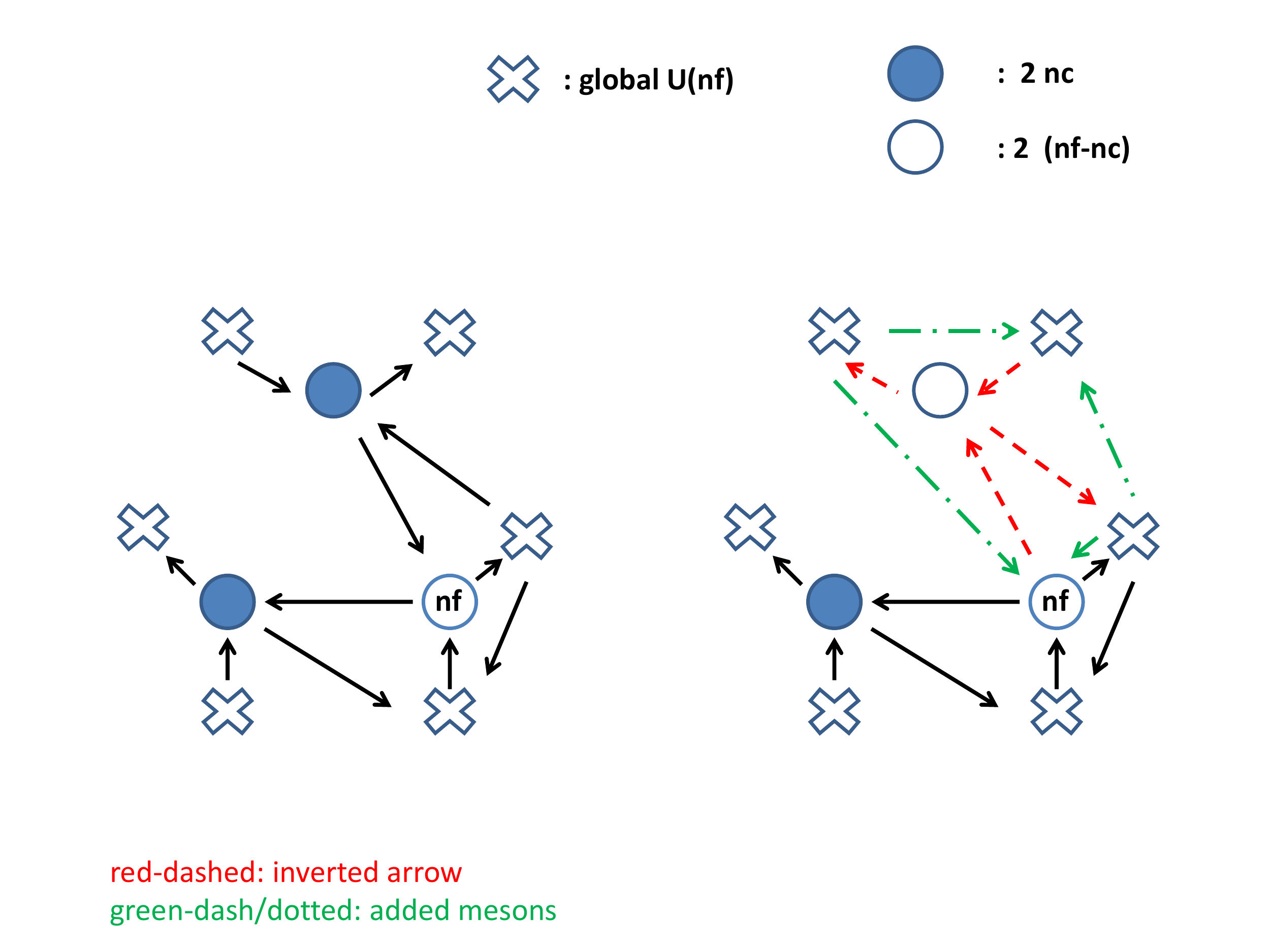}}
\caption{
Second step in the dualization for $k=3$. The algorithm is applied to the second node. Reverted arrows and added mesons are indicated as well as the dualized gauge group. The new quiver can be simplified by integrating out massive mesons, as indicated in the starting point for the next step in the following figure.
\label{k3two}
}
\end{center}
\end{figure}

\begin{figure}[!htbp]
\begin{center}
\scalebox{0.35}{
\includegraphics{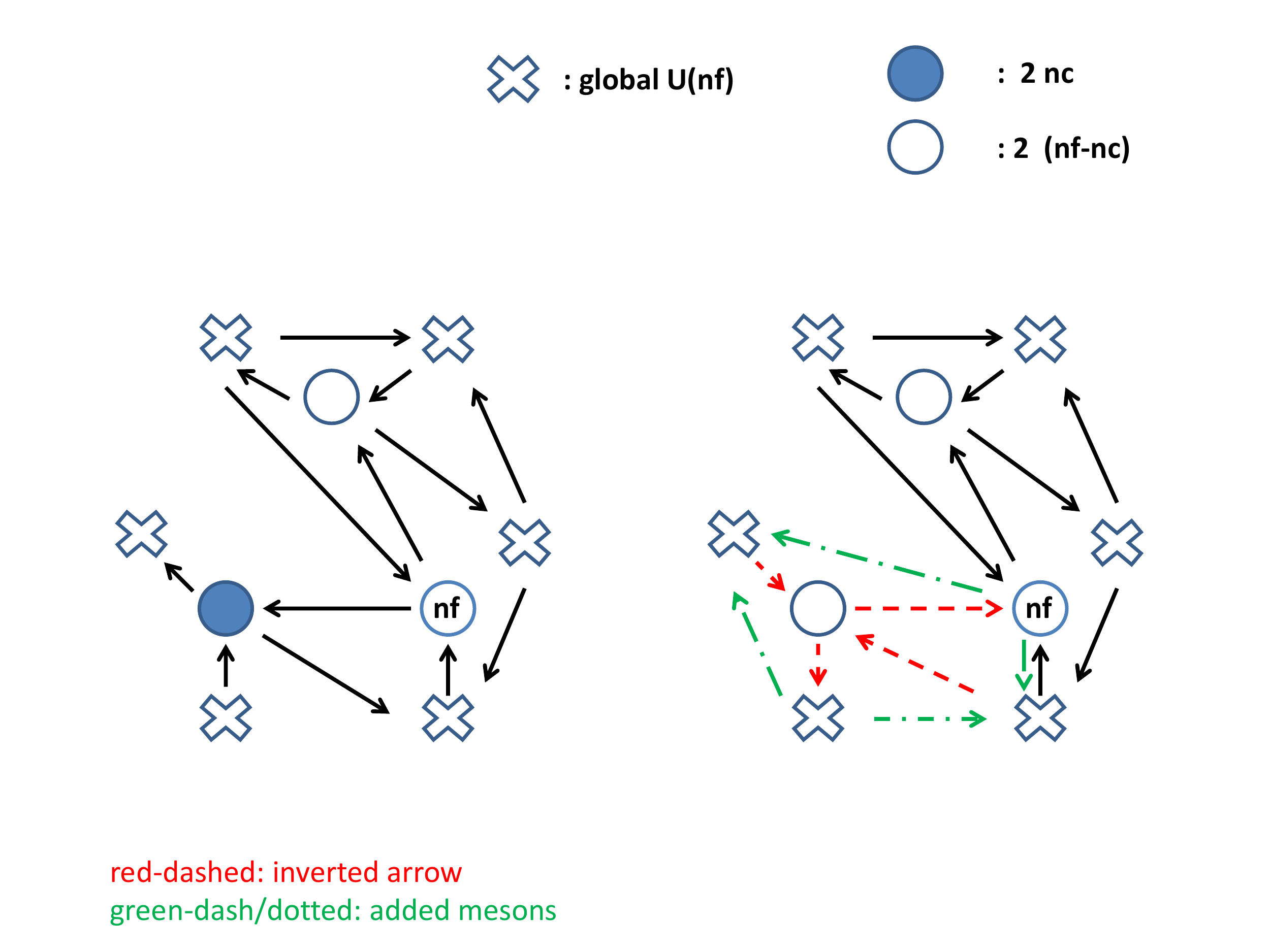}}
\caption{
 Third step in the dualization for $k=3$. The algorithm is applied to the third node. Reverted arrows and added mesons are indicated as well as the dualized gauge group. The new quiver can be simplified by integrating out massive mesons, as indicated in the starting point for the next step in the following figure.
\label{k3three}
}
\end{center}
\end{figure}

\begin{figure}[!htbp]
\begin{center}
\scalebox{0.35}{
\includegraphics{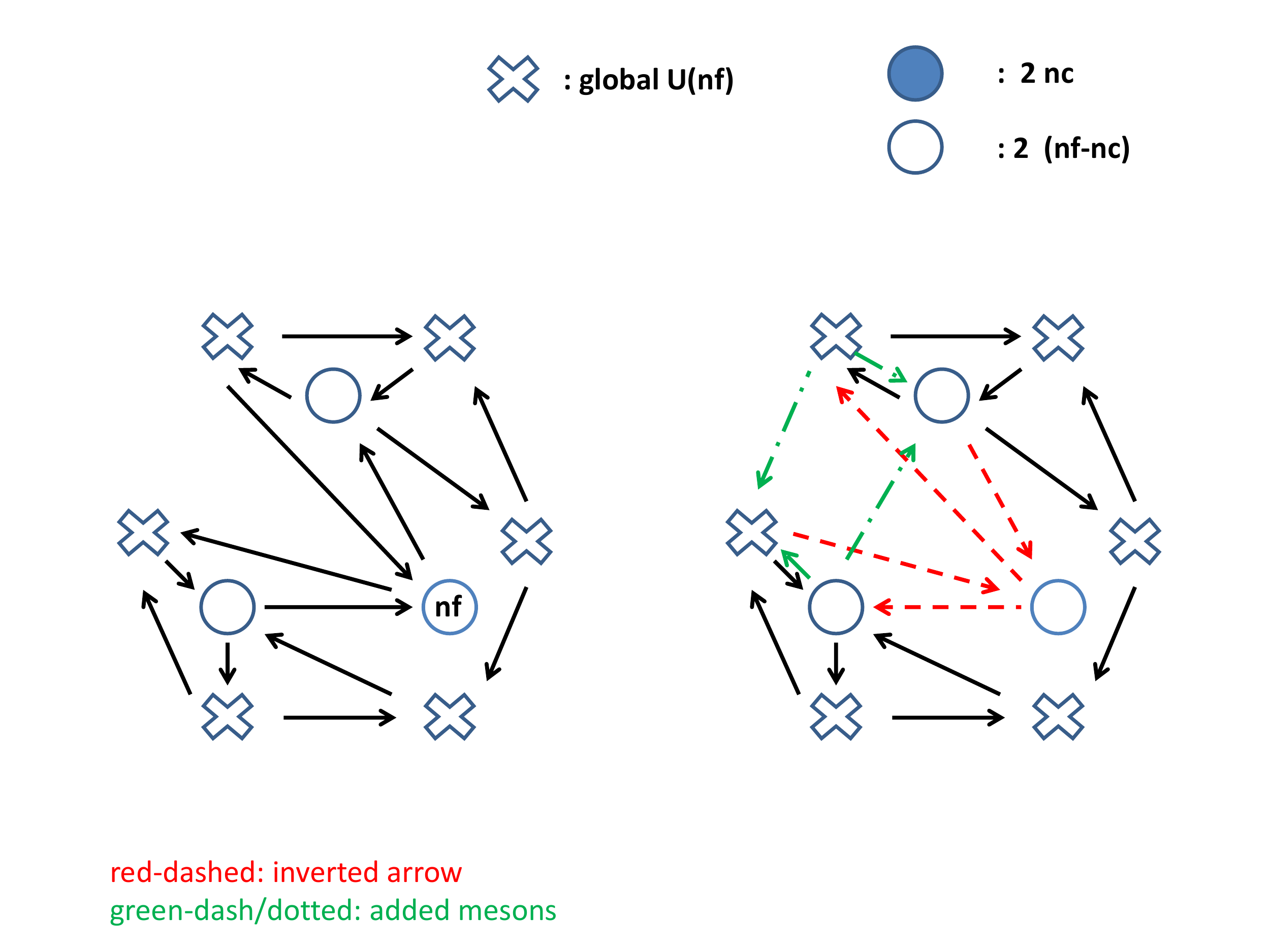}}
\caption{
 Fourth step in the dualization for $k=3$. The algorithm is applied to the first node. Reverted arrows and added mesons are indicated as well as the dualized gauge group. The new quiver can be simplified by integrating out massive mesons, as indicated in the next figure.
\label{k3four}
}
\end{center}
\end{figure}

\begin{figure}[!htbp]
\begin{center}
\scalebox{0.35}{
\includegraphics{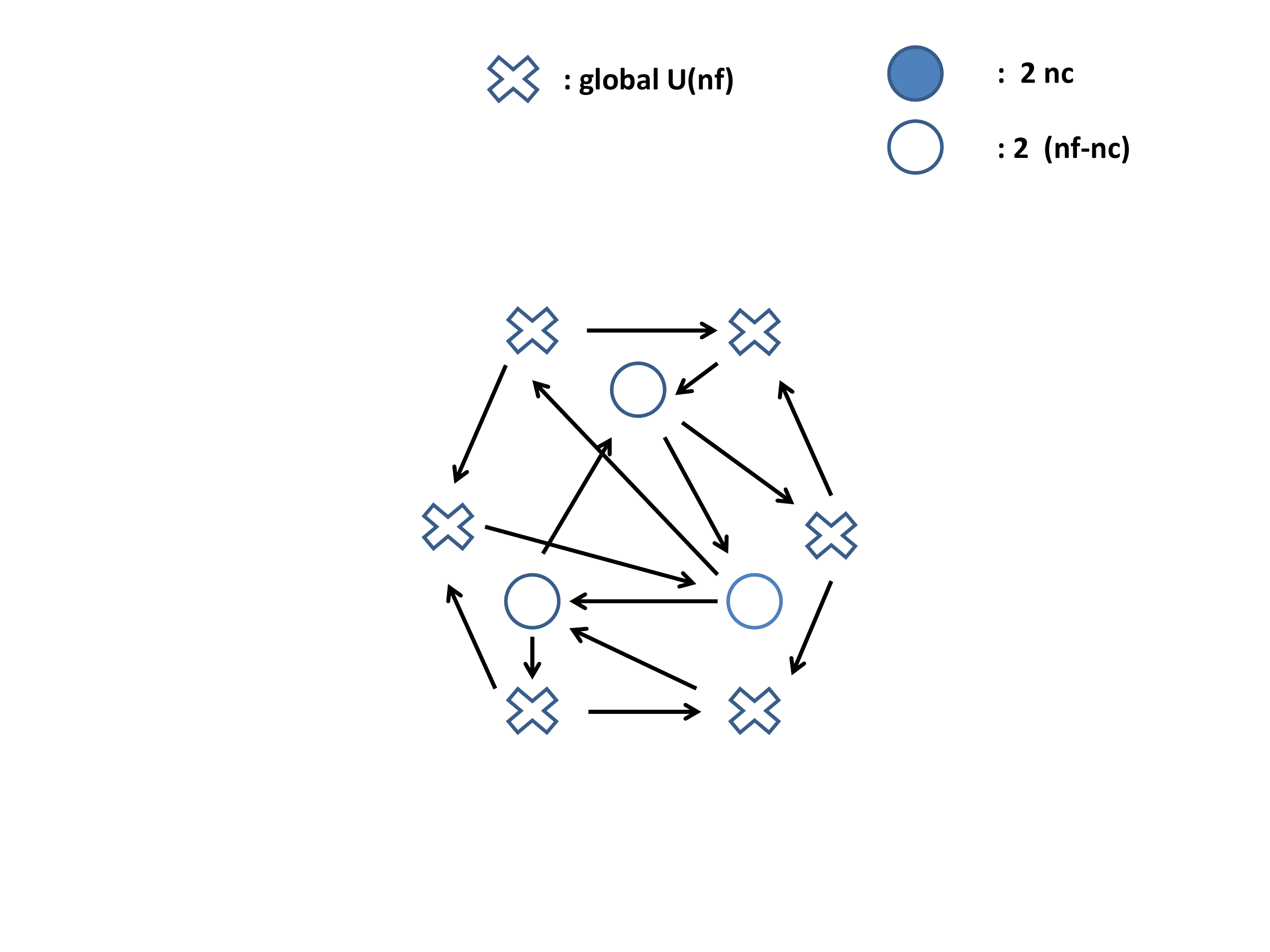}}
\caption{
Final quiver after implementing the dualization procedure for $k=3$.
\label{k3five}
}
\end{center}
\end{figure}

For a single node, dualizing involves a 5 step procedure:
\begin{enumerate}
\item Count the number of all ingoing arrows weighted by the number of colors/flavors associated with the node they come from. Equivalently one can count the outgoing arrows, they have to agree by anomaly cancelation. This gives the total number of flavors $N_f^{eff}$ for that node.
\item Dualize the gauge group on the node by replacing $N_c$ on that node with $N_f^{eff}-N_c$.
\item Reverse the orientation of all incoming and outgoing arrows associated with this node. This implements the fact that the dual quarks transform in the conjugate representation of the original quarks under the global flavor symmetry.
\item For every pair of an incoming and outgoing quark add a meson field that corresponds to the vector sum of the two. The orientation of the meson corresponds to the orientation the arrows had before they got inverted.
\item Whenever two arrows connect the same two nodes with opposite orientation, remove them from the quiver. As long as all superpotential terms associated with closed loops in the quiver had been added in the original theory, the $M q \tilde{q}$ superpotentials generated under dualization will generate a mass term for such a pair and it can be integrated out.
\end{enumerate}
The implementation of this algorithm for $k=3$ is illustrated graphically in figures (\ref{k3one}) - (\ref{k3five}). If we dualize the first node, then the second node, then the third node and then the first node one more time, we reach the desired dual theory.

This algorithm can easily be implemented in Mathematica. We find that if we dualize all $k$ nodes in order along the quiver, we reach a quiver where the first $k-2$ nodes now have gauge groups $U(n_f)$, $U(2 n_f)$, $\ldots$, $U((k-2) n_f)$. The last two nodes get dualized into the desired $U((k-1)(n_f - n_c))$ gauge group. Redualizing the first $k-2$ nodes ends up turning node $k-2$ into the desired dual. Redualizing the first $k-3$ nodes turns one more node into the correct dual gauge group. Repeating this process until every single mode has reached the dual takes a total of
\begin{equation}
N_{steps} = k + \sum_{I=1}^{k-2} (k-I-1) = \frac{k^2 - k + 2}{2}
\end{equation}
steps. Of course we can not exclude the possibility that there is a way to reach the dual gauge group in fewer steps than we indicated here, but we strongly suspect that one will always need order $k^2$ duality steps. Keeping track of the flavor content as well as the gauge singlet mesons, one sees the dual quiver emerge as expected. As in the $k=3$ example illustrated in figures (\ref{k3one}) - (\ref{k3five}) the flavor groups are ``rotated" with respect to the colored nodes along the quiver. But every color group has exactly one incoming and one outgoing flavor arrow. As expected, mesons connect every global quark flavor group (that is flavor groups with arrows from the flavor group to a color group) with every global anti-quark flavor group (that is flavor groups with arrows to the flavor group from a color group). We have explicitly verified this structure up to $k=100$.

In conclusion we find that indeed the orbifolded theories are dual to each other, so, for a given theory, the orbifold of the Seiberg dual is the same as the dual of the Seiberg orbifold if the orbifolded theory is supersymmetric.

\section*{Acknowledgements}
The authors would like to thank O.~Aharony, O.~Bergman, K.~Jensen and H.~Shimada for stimulating discussions and comments. The work of M.~H. is supported from Postdoctoral Fellowship for Research Abroad by Japan Society for the Promotion of Science. This work was supported in part by the U.S. Department of Energy under grant DE-FG02-96ER40956. C.H. was supported in part by the Israel Science Foundation (grant number 1468/06).

\end{document}